\documentclass[english]{IEEEtran}
\usepackage[T1]{fontenc}
\usepackage[latin9]{inputenc}
\usepackage{color}
\usepackage{amsthm}
\usepackage{amsmath}
\usepackage{amssymb}
\usepackage{graphicx}
\usepackage{esint}

\makeatletter

\providecommand{\tabularnewline}{\\}

\theoremstyle{plain}

\theoremstyle{plain}
\newtheorem{prop}{\protect\propositionname}
\theoremstyle{definition}
\newtheorem{defn}{\protect\definitionname}
\theoremstyle{remark}
\newtheorem{rem}{\protect\remarkname}
\theoremstyle{plain}
\newtheorem{lem}{\protect\lemmaname}

\usepackage{babel}
\usepackage{babel}
\usepackage{cite}

\providecommand{\propositionname}{Proposition}
\providecommand{\remarkname}{Remark}
\providecommand{\theoremname}{Theorem}

\addto\captionsenglish{}

\makeatother

\usepackage{babel}
\providecommand{\definitionname}{Definition}
\providecommand{\lemmaname}{Lemma}
\providecommand{\propositionname}{Proposition}
\providecommand{\remarkname}{Remark}
\providecommand{\theoremname}{Theorem}

\begin{document}

\title{\textbf{Generalized Labeled Multi-Bernoulli Approximation of Multi-Object
Densities}}

\author{Francesco Papi, Ba-Ngu Vo, Ba-Tuong Vo, Claudio Fantacci, and Michael
Beard %
\thanks{Acknowledgement: This work is supported by the Australian Research
Council under schemes DP130104404 and DE120102388.%
} %
\thanks{Francesco Papi, Ba-Ngu Vo, and Ba-Tuong Vo are with the Department
of Electrical and Computer Engineering, Curtin University, Bentley,
WA 6102, Australia. E-mail: \{francesco.papi, ba-tuong, ba-ngu.vo\}@curtin.edu.au%
} %
\thanks{Claudio Fantacci is with the Dipartimento di Ingegneria dell'Informazione,Università
di Firenze, Florence 50139, Italy. E-mail: claudio.fantacci@unifi.it%
} %
\thanks{Michael Beard is with Maritime Division, Defence Science and Technology
Organisation, Rockingham, WA 6958, Australia. E-mail: michael.beard@dsto.defence.gov.au%
} }
\maketitle
\begin{abstract}
In multi-object\textcolor{green}{{} }\textcolor{black}{inference}, the
multi-object probability density captures the uncertainty in the number
and the states of the objects as well as the statistical dependence
between the objects. Exact computation of the multi-object density
is generally intractable and tractable implementations usually require
statistical independence assumptions between objects. In this paper
we propose a tractable multi-object density approximation that can
capture statistical dependence between objects. In particular, we
derive a tractable Generalized Labeled Multi-Bernoulli (GLMB) density
that matches the cardinality distribution and the first moment of
the labeled multi-object distribution of interest. It is also shown
that the proposed approximation \textcolor{black}{minimizes the Kullback-Leibler
divergence} over a special tractable class of GLMB densities. Based
on the proposed GLMB approximation we further demonstrate a tractable
multi-object tracking algorithm for generic measurement models. Simulation
results for a multi-object Track-Before-Detect example using radar
measurements in low signal-to-noise ratio (SNR) scenarios verify the
applicability of the proposed approach.
\end{abstract}


\section{Introduction}

\IEEEPARstart{I}{n} multi-object inference the objective is the estimation
of an unknown number of objects and their individual states from noisy
observations. Multi-object estimation is a core problem in spatial
statistics \cite{Cressie91,Diggle03}, and multi-target tracking \cite{Mahler2007,Mahler2014},
spanning a diverse range of applications. Important applications of
spatial statistics include agriculture/forestry \cite{DrR97,Lundrudemo00,MollerWaage06},
epidemiology/public health \cite{Cressie91,Diggle03,WallerGotway04},
communications networks \cite{Baccelli,Haenggi,Haenggi09}, while
applications of multi-target tracking include radar/sonar \cite{Blackman1999,Bla04,Bar88},
computer vision \cite{BaddeleyVanLeishout92,Maggio2008,HVVS_PR_12,HVV_TSP_13},
autonomous vehicles \cite{Mullane2011,Lundquist2001,Lee2013,AVMM14},
automotive safety \cite{Battistelli,Meisner} and sensor networks
\cite{Zhang11,Lee12,Battistelli2013,Uney13}. \textcolor{black}{The
multi-object probability density is fundamental in multi-object estimation
because it captures the uncertainty in the number and the states of
the objects as well as the statistical dependence between the objects.
Statistical dependence between objects transpires via the data when
we consider the multi-object posterior density, or from the interactions
between objects as in Markov point processes \cite{RipleyKelly77,VanLieshout00}
or determinantal point processes \cite{Macchi75,Soshnikov00,Hough09}.}

\textcolor{black}{Computing the multi-object density is generally
intractable and approximations are necessary. Tractable multi-object
densities usually assume statistical independence between the objects.
For example, the Probability Hypothesis Density (PHD) \cite{MahlerPHD2},
Cardinalized PHD (CPHD) \cite{MahlerCPHD2007}, and multi-Bernoulli
filters \cite{VVC09}, are derived from multi-object densities in
which objects are statistically independent. On the other hand, multi-object
tracking approaches such as Multiple Hypotheses Tracking (MHT) \cite{Reid79mht,Kurien1990mht,Bla04}
and Joint Probabilistic Data Association (JPDA) \cite{Bar88} are
capable of modeling the statistical dependence between objects. However,
MHT does not have the notion of multi-object density while JPDA only
has the notion of multi-object density for a known number of objects.
A tractable family of multi-object densities that can capture the
statistical dependence between the objects is the recently proposed
Generalized Labeled Multi-Bernoulli (GLMB) family, which is conjugate
with respect to the standard measurement likelihood function\cite{BTV13,Vo2014}.}

The bulk of multi-object estimation algorithms in the literature,
including those discussed above, are designed for the so-called standard
measurement model, where data \textcolor{black}{has been preprocessed
into point measurements or detections \cite{Blackman1999,VVC09,Bla04,Bar88,MahlerCPHD2007}.
For a generic measurement model the GLMB density is not necessarily
a conjugate prior, i.e. the multi-object posterior density is not
a GLMB. This is the case in Track-Before-Detect (TBD) \cite{Boe01,Salmond2001,Ristic2004book,DaveyTBDbook,Pap13,PBTBV13},
tracking with superpositional measurements \cite{Nan13,MahSCPHD},
merged measurements \cite{Beard14}, and video measurements \cite{BTV10},\cite{HVVS_PR_12}.
In general, the multi-object density is numerically intractable in
applications involving non-standard measurement models. A simple strategy
that drastically reduces the numerical complexity is to approximate
the measurement likelihood by a separable likelihood \cite{BTV10}
for which Poisson, independently and identically distributed
(IID) cluster, multi-Bernoulli and GLMB densities are conjugate.
While this approximation can facilitate a trade off between tractability
and performance, biased estimates typically arise when the separable
assumption is violated.}

\textcolor{black}{Inspired by Mahler's IID cluster approximation in
the CPHD filter \cite{MahlerCPHD2007}, in this paper we consider
the approximation of a general labeled RFS density using a special
tractable class of GLMBs. In particular, we derive from this class
of GLMBs, an approximation to any labeled RFS density which preserves
the cardinality distribution and the first moment. It is also established
that our approximation minimizes the Kullback-Leibler divergence (KLD)
over this class of GLMB densities. This approximation is then applied
to develop an efficient multi-object tracking filter for a generic
measurement model. As an example application, we consider a radar
multi-object TBD problem with low signal-to-noise ratio (SNR) and
closely spaced targets. Simulation results verify that the proposed
approximation yields effective tracking performance in challenging
scenarios.}

\textcolor{black}{The paper is structured as follows: in Section II
we recall some definitions and results for Labeled random finite sets
(RFSs) and GLMB densities. In Section III we propose the G}LMB approximation
to multi-object distributions via cardinality, first moment matching
and KLD minimization. In Section IV we describe the application of
our result to multi-object tracking problems with non-standard measurement
models. Simulation results for challenging, low SNR, multi-target
TBD in radar scenarios are shown in Section V. Conclusions and future
research directions are reported in Section VI.

\section{Background}

\textcolor{black}{This section briefly presents background material
on multi-object filtering and labeled RFS which form the basis for
the formulation of our multi-object estimation problem.}

\subsection{\textcolor{black}{Labeled RFS}}

\textcolor{black}{An RFS on a space ${\mathcal{X}}$ is simply a random
variable taking values in ${\mathcal{F}}({\mathcal{X}})$, the space
of all finite subsets of ${\mathcal{X}}$. The space ${\mathcal{F}}({\mathcal{X}})$
does not inherit the Euclidean notion of integration and density.
Nonetheless, Mahler's Finite Set Statistics (FISST) provides powerful
yet practical mathematical tools for dealing with RFSs \cite{Goodman1997,MahlerPHD2,Mahler2007}
based on a notion of integration/density that is consistent with measure theory \cite{VSD05}.}

\textcolor{black}{A labeled RFS is an RFS whose elements are assigned
unique distinct labels \cite{BTV13}.
In this model, the single-object state space ${\mathcal{X}}$ is the
Cartesian product ${\mathbb{X}}{\mathcal{\times}}{\mathbb{L}}$, where
${\mathbb{X}}$ is the kinematic/feature space and ${\mathbb{L}}$
is the (discrete) label space. Let ${\mathcal{L}}:{\mathbb{X}}{\mathcal{\times}}{\mathbb{L}}\rightarrow{\mathbb{L}}$
be the projection ${\mathcal{L}}((x,\ell))=\ell$. A finite subset set ${\mathbf{X}}$
of ${\mathbb{X}}{\mathcal{\times}}{\mathbb{L}}$ has distinct labels
if ${\mathbf{X}}$ and its labels $\mathcal{L}(\mathbf{X})\triangleq\{\ell:(x,\ell)\in{\mathbf{X}}\}$
have the same cardinality. An RFS on ${\mathbb{X}}{\mathcal{\times}}{\mathbb{L}}$
with distinct labels is called a }\textcolor{black}{\emph{labeled
RFS}}\textcolor{black}{{} \cite{BTV13}.}

For the rest of the paper, we use the standard inner product notation
$\left\langle f,g\right\rangle \triangleq\int f(x)g(x)dx$, and multi-object
exponential notation $h^{X}\triangleq\prod_{_{x\in X}}h(x)$, where
$h$ is a real-valued function, with $h^{\emptyset}=1$ by convention.
We denote a generalization of the Kroneker delta and the inclusion
function which take arbitrary arguments such as sets, vectors, etc,
by
\begin{eqnarray*}
\delta_{Y}(X) & \triangleq & \left\{ \begin{array}{l}
1,{\text{ if }}X=Y\\
0,{\text{ otherwise}}
\end{array}\right.\\
1_{Y}(X) & \triangleq & \left\{ \begin{array}{l}
1,{\text{ if }}X\subseteq Y\\
0,{\text{ otherwise}}
\end{array}\right.
\end{eqnarray*}
We also write $1_{Y}(x)$ in place of $1_{Y}(\{x\})$ when $X=\{x\}$.
Single-object states are represented by lowercase letters, e.g. $x$,
${\mathbf{x}}$, while multi-object states are represented by uppercase
letters, e.g. $X$, ${\mathbf{X}}$, symbols for labeled states and
their distributions are bolded to distinguish them from unlabeled ones,
e.g. ${\mathbf{x}}$, ${\mathbf{X}}$, ${\mathbf{\pi}}$, etc, spaces
are represented by blackboard bold e.g. ${\mathbb{X}}$, ${\mathbb{Z}}$,
${\mathbb{L}}$, etc. \textcolor{black}{The integral of a function
$f$ on ${\mathbb{X}}{\mathcal{\times}}{\mathbb{L}}$ is given by
\[
\int f({\bf {\bf x}})d{\bf x}=\sum_{\ell\in{\mathbb{L}}}\int f(x,\ell)dx.
\]
Two important statistics of an RFS relevant to this paper are the
cardinality distribution $\rho(\cdot)$ and the PHD $v(\cdot)$ \cite{Mahler2007}:
\begin{align}
\rho(n) & =\frac{1}{n!}\int\boldsymbol{\pi}(\{\mathbf{x}_{1},...,\mathbf{x}_{n}\})d(\mathbf{x}_{1},...,\mathbf{x}_{n})\label{eq:CARD}\\
v(x,\ell) & =\int\boldsymbol{\pi}({\left\{ (x,\ell)\right\} \cup\mathbf{X}})\delta\mathbf{X}\label{eq:PHD}
\end{align}
where the integral is a }\textcolor{black}{\emph{set integral}}\textcolor{black}{{}
defined for any function $f$ on ${\mathcal{F}}({\mathcal{X}})$ by
\[
\int f({\bf {\bf X}})\delta{\bf X}=\sum_{i=0}^{\infty}\frac{1}{i!}\int f(\{\mathbf{x}_{1},...,\mathbf{x}_{i}\})d({\mathbf{x}}_{1},...,\mathbf{x}_{i}).
\]
The PHD in (\ref{eq:PHD}) and the unlabeled PHD in \cite{BTV13},
i.e. the PHD of the unlabeled version, are related by $v(x)=\sum_{\ell\in\mathbb{L}}v(x,\ell)$. Hence, $v(\cdot,\ell)$ can be interpreted as the contribution from
label $\ell$ to the unlabeled PHD.}

\subsection{Generalized Labeled Multi-Bernoulli}

\textcolor{black}{An important class of labeled RFS is the generalized
labeled multi-Bernoulli (GLMB) family \cite{BTV13}, which forms the
basis of an analytic solution to the Bayes multi-object filter \cite{Vo2014}.
Under the standard multi-object likelihood, the GLMB is a conjugate
prior, which is also closed under the Chapman-Kolmogorov equation
\cite{BTV13}. Thus if initial prior is a GLMB density, then the multi-object
prediction and posterior densities at all subsequent times are also
GLMB densities.}

A GLMB is an RFS of ${\mathbb{X}}{\mathcal{\times}}{\mathbb{L}}$
distributed according to
\begin{equation}
\mathbf{\boldsymbol{\pi}}(\mathbf{X})=\Delta(\mathbf{X})\sum_{c\in\mathbb{C}}w^{(c)}(\mathcal{L}(\mathbf{X}))\left[p^{(c)}\right]^{\mathbf{X}}\label{eq:GLMB}
\end{equation}
where $\Delta({\mathbf{X}})\triangleq$$\delta_{|{\mathbf{X}}|}(|{\mathcal{L}}({\mathbf{X}})|)$
denotes the \emph{distinct label indicator}, ${\mathbb{C}}$\textcolor{black}{{} is a discrete index set, and
$w^{(c)}$, $p^{(c)}$ satisfy:}
\begin{eqnarray}
\sum_{L\subseteq\mathbb{L}}\sum_{c\in\mathbb{C}}w^{(c)}(L) & = & 1,\label{eq:GLMBsumw}\\
\int p^{(c)}(x,\ell)dx & = & 1.
\end{eqnarray}
The GLMB density (\ref{eq:GLMB}) can be interpreted as a mixture
of multi-object exponentials. Each term in (\ref{eq:GLMB}) consists
of a weight $w^{(c)}({\mathcal{L}}({\mathbf{X}}))$ that depends only
on the labels of ${\mathbf{X}}$, and a multi-object exponential $\left[p^{(c)}\right]^{{\mathbf{X}}}$
that depends on the labels and kinematics/features of ${\mathbf{X}}$.

The cardinality distribution and PHD of a GLMB are, respectively, given
by \cite{BTV13}
\begin{align}
\rho(n) & =\sum_{c\in{\mathbb{C}}}\sum_{L\subseteq{\mathbb{L}}}\delta_{n}(|L|)w^{(c)}(L),\label{eq:GLMBCard}\\
v(x,\ell) & =\sum_{c\in{\mathbb{C}}}p^{(c)}(x,\ell)\sum_{L\subseteq{\mathbb{L}}}1_{L}(\ell)w^{(c)}(L).\label{eq:lGLMBLabeledPHD}
\end{align}


\textcolor{black}{A Labeled Multi-Bernoulli (LMB) density is a special
case of the GLMB density with one term (in which case the superscript $(c)$
is not needed) and a specific form for the only weight $w(\cdot)$
\cite{BTV13,Reuter2014}:
\begin{eqnarray}
w(L) & = & \prod\limits _{\ell\in{\mathbb{M}}}\left(1-r^{(\ell)}\right)\prod\limits _{\ell\in L}\frac{1_{{\mathbb{M}}}(\ell)r^{(\ell)}}{1-r^{(\ell)}},
\end{eqnarray}
where $r^{(\ell)}$ for $\ell\in{\mathbb{M\subseteq L}}$ represents
the existence probability of track $\ell$, and $p(\cdot,\ell)$ is
the probability density of the kinematic state of track $\ell$ conditional
upon existence \cite{BTV13}.} Note that the LMB density can always be factored into a product of
terms over the elements of $\mathbf{X}$. The LMB density can thus
be interpreted as comprising multiple independent tracks. The LMB
density is in fact the basis of the LMB filter, a principled and efficient
approximation of the Bayes multi-object tracking filter, which is
highly parallelizable and capable of tracking large numbers of targets
\cite{Reuter2014,LMB2014manyTargets}.

\section{Multi-Object Estimation with GLMBs{\normalsize{}\label{sec:GLMBapprox}}}

In this section we discuss the multi-object estimation problem with
GLMBs. In particular, in subsection \ref{sub:SepLK} we present a
simple approximation through a \emph{separable likelihood }function
which exploits the conjugacy of the GLMB distributions, while in subsection
\ref{sub:NonSepLK} we propose a more principled approach for approximating
a general labeled RFS density with a special form GLMB that matches
both the PHD and cardinality distribution.

\subsection{Conjugacy with respect to Separable Likelihoods{\normalsize{}\label{sub:SepLK}}}

\textcolor{black}{A separable multi-object likelihood }of the state
$\mathbf{X}$ given the measurement $z$ \textcolor{black}{is one
of the form \cite{BTV10}:
\begin{eqnarray}
g(z|{\mathbf{X}}) & \propto & \gamma_{z}^{{\mathbf{X}}}=\prod_{\mathbf{x}\in\mathbf{X}}\gamma_{z}(\mathbf{x})\label{eq:sep_lk}
\end{eqnarray}
where $\gamma_{z}(\cdot)$ is a non-negative function defined on $\mathbb{\mathbb{X}}$. }

\textcolor{black}{It was shown in \cite{BTV10} that Poisson, IID
cluster and multi-Bernoulli densities are conjugate with respect to
separable multi-object likelihood functions. Moreover, this conjugacy
is easily extented to the family of GLMBs. }
\begin{prop}
\textcolor{black}{If the multi-object prior density $\boldsymbol{\pi}$
is a GLMB of the form (\ref{eq:GLMB}) and the multi-object likelihood
is separable of the form (\ref{eq:sep_lk}), then the multi-object
posterior density is a GLMB of the form:
\begin{equation}
{\mathbf{\boldsymbol{\pi}}}({\mathbf{X}}|z)\propto\Delta({\mathbf{X}})\sum\limits _{c\in{\mathbb{C}}}w_{z}^{(c)}({\mathcal{L}}({\mathbf{X}}))\left[p^{(c)}(\cdot|z)\right]^{{\mathbf{X}}}\label{eq:PropConj0-1}
\end{equation}
where
\begin{eqnarray}
w_{z}^{(c)}(L) & = & \left[\eta_{z}\right]^{L}w^{(c)}(L)\label{eq:PropConj1-1}\\
p^{(c)}(x,\ell|z) & = & p^{(c)}(x,\ell)\gamma_{z}(x,\ell)/\eta_{z}(\ell)\label{eq:PropConj2-1}\\
\eta_{z}(\ell) & = & \left\langle p^{(c)}(\cdot,\ell),\gamma_{z}(\cdot,\ell)\right\rangle \label{eq:PropConj3-1}
\end{eqnarray}
\vspace{-5mm}}\end{prop}
\emph{Proof:}
\textcolor{black}{
\begin{eqnarray*}
\boldsymbol{\pi}({\mathbf{X}}|z) & \propto & \gamma_{z}^{{\mathbf{X}}}\boldsymbol{\pi}({\mathbf{X}})\\
 & = & \Delta({\mathbf{X}})\sum\limits _{c\in{\mathbb{C}}}w^{(c)}({\mathcal{L}}({\mathbf{X}}))\gamma_{z}^{{\mathbf{X}}}[p^{(c)}]^{{\mathbf{X}}}\\
 & = & \Delta({\mathbf{X}})\sum\limits _{c\in{\mathbb{C}}}w^{(c)}({\mathcal{L}}({\mathbf{X}}))\left[\eta_{z}\right]^{{\mathcal{L}}({\mathbf{X}})}\frac{\left[\gamma_{z}p^{(c)}\right]^{\mathbf{X}}}{\left[\eta_{z}\right]^{{\mathcal{L}}({\mathbf{X}})}}\\
 & = & \Delta({\mathbf{X}})\sum\limits _{c\in{\mathbb{C}}}w_{z}^{(c)}({\mathcal{L}}({\mathbf{X}}))\left[p^{(c)}(\cdot|z)\right]^{{\mathbf{X}}}. \ \ \ \ \ \ \ \ \  \blacksquare 
\end{eqnarray*}
}

In general, the true multi-object likelihood is not separable, however
the separable likelihood assumption can be a reasonable approximation
if the objects do not overlap in the measurement space \cite{BTV10}.

\subsection{Labeled RFS Density Approximation{\normalsize{} \label{sub:NonSepLK}}}

In this subsection we propose a tractable GLMB
density approximation to an arbitrary labeled multi-object density
$\boldsymbol{\pi}$. Tractable GLMB densities are numerically evaluated
via the so-called $\delta$-GLMB form which involves explicit enumeration of the label
sets (for more details see \cite{BTV13,Vo2014}). Since there is no
general information on the form of $\boldsymbol{\pi}$, a natural
choice is the class of $\delta$-GLMBs of the form
\begin{equation}
\mathbf{\boldsymbol{\bar{\pi}}}(\mathbf{X})=\Delta({\mathbf{X}})\sum_{L\in{\mathcal{F}}({\mathbb{L}})}\bar{w}^{(L)}\delta_{L}({\mathcal{L}}(\mathbf{X}))\left[\bar{p}^{(L)}\right]^{\mathbf{X}}\label{eq:ApproximateGLMB-1-1}
\end{equation}
where each $\bar{p}^{(L)}(\cdot,\ell)$ is a density on $\mathbb{X}$, and
each weight $\bar{w}^{(L)}$ is non-negative such that $\sum_{L\subseteq\mathbb{L}}w^{(L)}=1$. 
 It follows from (\ref{eq:GLMBCard}) and (\ref{eq:lGLMBLabeledPHD}) that the cardinality distribution and PHD of (\ref{eq:ApproximateGLMB-1-1}) are
given, respectively, by
\begin{align}
\bar{\rho}(n) & =\sum_{L\subseteq{\mathbb{L}}}\delta_{n}(|L|)\bar{w}^{(L)},\label{eq:d-GLMBCard}\\
\bar{v}(x,\ell) & =\sum_{L\subseteq{\mathbb{L}}}1_{L}(\ell)\bar{w}^{(L)}\bar{p}^{(L)}(x,\ell). \label{eq:d-lGLMBLabeledPHD}
\end{align}
Note that such $\delta$-GLMB is completely characterised by the parameter
set $\{(\bar{w}^{(L)},\bar{p}^{(L)})\}_{L\in{\mathcal{F}}({\mathbb{L}})}$. Our objective is to seek a density, via its parameter set, from this
class of $\delta$-GLMBs, which matches the PHD and cardinality distribution
of $\boldsymbol{\pi}$.

The strategy of matching the PHD and cardinality
distribution is inspired by Mahler's IID cluster approximation in
the CPHD filter \cite{MahlerCPHD2007}, which has proven to be very
effective in practice \cite{Mahler2014,Georgescu2009,Svensson09}.
While our result is used to develop a multi-object tracking algorithm
in the next section, it is not necessarily restricted to tracking
applications, and can be used in more general multi-object estimation problems.

\textcolor{black}{Our result follows from the following representation
for labeled RFS.}
\begin{defn}
\textcolor{black}{Given a labeled multi-object density $\boldsymbol{\pi}$
on ${\mathcal{F}}({\mathbb{X}}{\mathcal{\times}}{\mathbb{L}})$, and
any positive integer $n$, we define the }\textcolor{black}{\emph{joint
existence probability}}\textcolor{black}{{} of the label set $\{\ell_{1},...,\ell_{n}\}$
by
\begin{eqnarray}
w(\{\ell_{1},...,\ell_{n}\})\ \ \ \ \ \ \ \ \ \ \ \ \ \ \ \ \ \ \ \ \ \ \ \!\ \ \ \ \ \ \ \ \ \ \ \ \ \ \ \ \ \ \ \ \ \ \ \nonumber \\
\triangleq\int\boldsymbol{\pi}(\{(x_{1},\ell_{1}),...,(x_{n},\ell_{n})\})d(x_{1},...,x_{n})\label{eq:GLMB1}
\end{eqnarray}
and the}\textcolor{black}{\emph{ joint probability density}}\textcolor{black}{{}
(on ${\mathbb{X}}^{n}$) }\textcolor{black}{\emph{of}}\textcolor{black}{{}
$x_{1},...,x_{n}$, }\textcolor{black}{\emph{conditional on their
corresponding labels}}\textcolor{black}{{} $\ell_{1},...,\ell_{n}$,
by
\begin{equation}
\!p(\{(x_{1},\ell_{1}),...,(x_{n},\ell_{n})\})\triangleq\frac{\boldsymbol{\pi}(\{(x_{1},\ell_{1}),...,(x_{n},\ell_{n})\})}{w(\{\ell_{1},...,\ell_{n}\})}\label{eq:GLMB2}
\end{equation}
For $n=0$, we define $\ensuremath{w(\emptyset)\triangleq\boldsymbol{\pi}(\emptyset)}$
and $\ensuremath{p(\emptyset)\triangleq1}$. It is implicit that $p({\mathbf{X}})$
is defined to be zero whenever $w({\mathcal{L}}({\mathbf{X}}))$ is
zero. Consequently, the labeled multi-object density can be expressed
as
\begin{equation}
\boldsymbol{\pi}({\mathbf{X}})=w({\mathcal{L}}({\mathbf{X}}))p({\mathbf{X}})\label{eq:GLMBjoint}
\end{equation}
\vspace{-5mm}}\end{defn}
\begin{rem}
\textcolor{black}{Note that $\sum_{L\in{\mathcal{F}}({\mathbb{L}})}w(L)=1$,
and since $\boldsymbol{\pi}$ is symmetric in its arguments it follows
from Lemma \ref{lem:symmetric} that $w(\cdot)$ is also symmetric
in $\ell_{1},...,\ell_{n}$. Hence $w(\cdot)$ is indeed a probability
distribution on ${\mathcal{F}}({\mathbb{L}})$.}\end{rem}
\begin{lem}
\textcolor{black}{\label{lem:symmetric}Let $f:(\mathbb{X}\times\mathbb{Y})^{n}\rightarrow\mathbb{R}$
be symmetric. Then $g:\mathbb{X}{}^{n}\rightarrow\mathbb{R}$ given
by
\[
g(x_{1},...,x_{n})=\int f((x_{1},y_{1}),...,(x_{n},y_{n}))d(y_{1},...,y_{n})
\]
is also symmetric on $\mathbb{X}^{n}$.}\end{lem}
\begin{IEEEproof}
\textcolor{black}{Let $\sigma$ be a permutation of $\{1,...,n\}$,
then
\begin{align*}
 & \!\!\!\!g(x_{\sigma(1)},...,x_{\sigma(n)})\\
 & =\int \! f((x_{\sigma(1)},y_{\sigma(1)}),...,(x_{\sigma(n)},y_{\sigma(n)}))d(y_{\sigma(1)},...,y_{\sigma(n)})\\
 & =\int \! f((x_{1},y_{1}),...,(x_{n},y_{n}))d(y_{\sigma(1)},...,y_{\sigma(n)})\\
 & =\int \! f((x_{1},y_{1}),...,(x_{n},y_{n}))d(y_{1},...,y_{n})
\end{align*}
where the last step follows from the fact that the order of integration
is interchangeable.}\end{IEEEproof}
\begin{prop}
\textcolor{black}{\label{prop:2}Given any labeled multi-object density
$\mathbf{\boldsymbol{\pi}}$, the $\delta$-GLMB density in the class
defined by (\ref{eq:ApproximateGLMB-1-1}) which preserves the cardinality
distribution and PHD of $\boldsymbol{\mathbb{\pi}}$, and minimizes
the Kullback-Leibler divergence from $\mathbf{\boldsymbol{\pi}}$,
is given by
\begin{equation}
\boldsymbol{\hat{\pi}}(\mathbf{X})=\Delta({\mathbf{X}})\sum_{I\in{\mathcal{F}}({\mathbb{L}})}\hat{w}^{(I)}\delta_{I}({\mathcal{L}}({\mathbf{X}}))\left[\hat{p}^{(I)}\right]^{{\mathbf{X}}}\label{eq:ApproximateGLMB}
\end{equation}
where
\begin{align}
 \!\! \hat{w}^{(I)}&=w(I),\label{eq:MarginalizeGeneral-1-1}\\
 \!\! \hat{p}^{(I)}(x,\ell)&=1_{I}(\ell)p_{I-\{\ell\}}(x,\ell),\label{eq:MarginalizeGeneral-2-1}\\
 \!\! p_{\{\ell_{1},...,\ell_{n}\}}(x,\ell)&=\nonumber \\
  &\!\!\!\!\!\!\!\!\!\!\!\!\!\!\!\!\!\!\!\!\!\! \int \! p(\left\{ (x,\ell),(x_{1},\ell_{1}),...,(x_{n},\ell_{n})\right\} )d\left(x_{1},...,x_{n}\right).\label{eq:MarginalizeGeneral-3-1}
\end{align}}\end{prop}
\vspace{-0.5cm}
\begin{rem}
Note from the definition of $\hat{p}^{(I)}(x,\ell)$ in (\ref{eq:MarginalizeGeneral-2-1})
that
\begin{eqnarray}
 &  & \!\!\!\!\!\!\!\!\!\!\!\!\! \hat{p}^{(\{\ell,\ell_{1},...,\ell_{n}\})}(x,\ell)\nonumber \\
 &  & \ =\int\!{p(}\{(x,\ell),(x_{1},\ell_{1}),...,(x_{n},\ell_{n})\}{)d}(x_{1},...,x_{n})\ \ \ \ \ \label{eq:pImarg}
\end{eqnarray}
Hence,  $\hat{p}^{(\{\ell_{1},...,\ell_{n}\})}(\cdot,\ell_{i})$, $i=1,...,n$, defined in (\ref{eq:MarginalizeGeneral-2-1}) are the marginals of the label-conditioned joint density $p(\{(\cdot,\ell_{1}),...,(\cdot,\ell_{n})\})$ of $\mathbf{\pi}$.

Proposition \ref{prop:2} states that replacing the label-conditioned joint densities, of a labeled multi-object density $\boldsymbol{{\pi}}$, by the products of their marginals yields a $\delta$-GLMB of the form (\ref{eq:ApproximateGLMB-1-1}), which  minimises the Kullback-Leibler divergence from $\boldsymbol{{\pi}}$, and matches its PHD and cardinality distribution.
\end{rem}

\begin{IEEEproof}
\textcolor{black}{Since $p_{\{\ell_{1},...,\ell_{n}\}}(x,\ell)$ is symmetric in $\ell_{1},...,\ell_{n}$,
via Lemma \ref{lem:symmetric}, $\hat{p}^{(I)}(x,\ell)$ is
indeed a function of the set $I$. The proof uses the fact (\ref{eq:ApproximateGLMB-1-1})
can be rewritten as $\mathbf{\boldsymbol{\bar{\pi}}}({\mathbf{X}})=\bar{w}({\mathcal{L}}({\mathbf{X}}))\bar{p}({\mathbf{X}})$
where
\begin{align*}
\bar{w}(L) & =\bar{w}^{(L)},\\
\bar{p}({\mathbf{X}}) & =\Delta({\mathbf{X}})\left[\bar{p}^{({\mathcal{L}}({\mathbf{X}}))}\right]^{\mathbf{X}}.
\end{align*}
To show that $\boldsymbol{\hat{\pi}}$ preserves the cardinality of
$\mathbf{\boldsymbol{\pi}}$, observe that the cardinality distribution
of any labeled RFS is completely determined by the joint existence
probabilities of the labels $w(\cdot)$, i.e.
\begin{align*}
\rho(n) & =\frac{1}{n!}\sum_{\left(\ell_{1},...,\ell_{n}\right)\in{\mathbb{L}}^{n}}\int w(\{\ell_{1},...,\ell_{n}\})\times\\
 & \ \ \ \ \ \ \ \ \ \ \ \ \ \ \ \ \ p(\{(x_{1},\ell_{1}),...,(x_{n},\ell_{n})\})d(x_{1},...,x_{n})\\
 & =\sum_{L\subseteq{\mathbb{L}}}\delta_{n}(|L|)w(L)
\end{align*}
Since both $\boldsymbol{\hat{\pi}}$ and $\mathbf{\boldsymbol{\pi}}$
have the same joint existence probabilities, i.e. $\hat{w}(L)=\hat{w}^{(L)}=w(L)$,
their cardinality distributions are the same.}

\textcolor{black}{To show that the PHDs of $\boldsymbol{\hat{\pi}}$
and $\mathbf{\boldsymbol{\pi}}$ are the same, note from (\ref{eq:d-lGLMBLabeledPHD}) that the PHD of $\boldsymbol{\hat{\pi}}$ can be expanded as 
\begin{flalign*}
  \hat{v}(x,\ell) &=\sum_{n=0}^{\infty}\frac{1}{n!}\sum_{\left(\ell_{1},...,\ell_{n}\right)\in{\mathbb{L}}^{n}}\hat{w}^{(\{\ell,\ell_{1},...,\ell_{n}\})}\hat{p}^{(\{\ell,\ell_{1},...,\ell_{n}\})}(x,\ell)\\
  &=\sum_{n=0}^{\infty}\frac{1}{n!}\sum_{\left(\ell_{1},...,\ell_{n}\right)\in{\mathbb{L}}^{n}}w(\{\ell,\ell_{1},...,\ell_{n}\})\times\\
 & \ \ \ \ \ \ \ \  \int p(\{(x,\ell),(x_{1},\ell_{1}),...,(x_{n},\ell_{n})\})d(x_{1},...,x_{n})
 \vspace{-0.2cm}
\end{flalign*}
where the last step follows by substituting (\ref{eq:MarginalizeGeneral-1-1}) and (\ref{eq:pImarg}). The right hand side of the above equation is the set integral
$\int\boldsymbol{\pi}({\left\{ \left(x,\ell\right)\right\} \cup\mathbf{X}})\delta\mathbf{X}$.
Hence $\hat{v}(x,\ell)=v(x,\ell)$.}

\textcolor{black}{The Kullback-Leibler divergence from $\mathbf{\boldsymbol{\pi}}$
and any $\delta$-GLMB of the form (\ref{eq:ApproximateGLMB-1-1})
is given by
\begin{flalign*}
\vspace{-0.1cm}
 & \mbox{\ensuremath{D_{KL\!}}(}{\mathbf{\boldsymbol{\pi}}};\mathbf{\boldsymbol{\bar{\pi}}})\\
 & =\int\log\left(\frac{w({\mathcal{L}}({\mathbf{X}}))p({\mathbf{X}})}{\bar{w}({\mathcal{L}}({\mathbf{X}}))\bar{p}({\mathbf{X}})}\right)w({\mathcal{L}}({\mathbf{X}}))p({\mathbf{X}})\delta\mathbf{X}\\
 & =\sum_{n=0}^{\infty}\frac{1}{n!}\sum_{\left(\ell_{1},...,\ell_{n}\right)\in{\mathbb{L}}^{n}}\log\left(\frac{w(\{\ell_{1},...,\ell_{n}\})}{\bar{w}(\{\ell_{1},...,\ell_{n}\})}\right)\times\\
 & \ \ \ \ \ \ \ \ w(\{\ell_{1},...,\ell_{n}\})\!\!\int \!\! p(\{(x_{1},\ell_{1}),...,(x_{n},\ell_{n})\})d(x_{1},...,x_{n})\\
 & +\sum_{n=0}^{\infty}\frac{1}{n!}\sum_{\left(\ell_{1},...,\ell_{n}\right)\in{\mathbb{L}}^{n}}\int\log\left(\frac{p(\{(x_{1},\ell_{1}),...,(x_{n},\ell_{n})\})}{\prod_{i=1}^{n}\bar{p}^{(\{\ell_{1},...,\ell_{n}\})}(x_{i},\ell_{i})}\right)\times\\
 & \ \ \ \ \ \ \ \ \ \ w(\{\ell_{1},...,\ell_{n}\})p(\{(x_{1},\ell_{1}),...,(x_{n},\ell_{n})\})d(x_{1},...,x_{n})
\end{flalign*}
Noting that $p(\{(\cdot,\ell_{1}),...,(\cdot,\ell_{n})\})$ integrates to 1, we have
\vspace{-0.1cm}
\begin{flalign*}
 & \mbox{\ensuremath{D_{KL}}(}\mathbf{\boldsymbol{\pi}};\mathbf{\boldsymbol{\bar{\pi}}})=\\
 & \ \ \ \ \ \ \ \ \ \ \ \  \mbox{\ensuremath{D_{KL\!}}}\left(w;\bar{w}\right)+\sum_{n=0}^{\infty}\frac{1}{n!}\sum_{\left(\ell_{1},...,\ell_{n}\right)\in{\mathbb{L}}^{n}}w(\{\ell_{1},...,\ell_{n}\})\times\\
 & \ \ \ \ \ \ \ \ \ \ \ \ \mbox{\ensuremath{D_{KL\!\!}}}\left(p(\{(\cdot,\ell_{1}),...,(\cdot,\ell_{n})\});\prod_{i=1}^{n}\bar{p}^{(\{\ell_{1},...,\ell_{n}\})}(\cdot,\ell_{i})\!\right)
 \vspace{-0.1cm}
\end{flalign*}
Setting ${\boldsymbol{\bar{\pi}}}=\boldsymbol{\hat{\pi}}$ we have $\mbox{\ensuremath{D_{KL}}}\left(w;\hat{w}\right)=0$ since $\hat{w}(I)=w(I)$.
Moreover, for each $n$ and each $\{\ell_{1},...,\ell_{n}\}$,
$\hat{p}^{(\{\ell_{1},...,\ell_{n}\})}(\cdot,\ell_{i})$, $i=1,...,n$,
are the marginals of $p(\{(\cdot,\ell_{1}),...,(\cdot,\ell_{n})\})$.
Hence, it follows from \cite{cardoso2003} that each
Kullback-Leibler divergence in the above sum is minimized. Therefore, $\mbox{\ensuremath{D_{KL}}(}{\mathbf{\boldsymbol{\pi}}};\boldsymbol{\hat{\pi}})$
is minimized over the class of $\delta$-GLMB
of the form (\ref{eq:ApproximateGLMB-1-1}).}
\end{IEEEproof}
\textcolor{black}{The cardinality and PHD matching strategy in the above Proposition
can be readily extended to the approximation of any labeled multi-object
density of the form
\begin{equation}
\boldsymbol{\pi}({\mathbf{X}})=\Delta({\mathbf{X}})\sum\limits _{c\in{\mathbb{C}}}w^{(c)}({\mathcal{L}}({\mathbf{X}}))p^{(c)}({\mathbf{X}})\label{eq:PropConj0-1-1}
\end{equation}
where the weights $w^{(c)}(\cdot)$ satisfy (\ref{eq:GLMBsumw}) and
\begin{eqnarray}
\int p^{(c)}(\{(x_{1},\ell_{1}),...,(x_{n},\ell_{n})\})d(x_{1},...,x_{n})=1
\vspace{-0.05cm}
\end{eqnarray}
by approximating each $p^{(c)}(\{(\cdot,\ell_{1}),...,(\cdot,\ell_{n})\})$ by the product of its marginals. This is a better approximation than directly applying Proposition \ref{prop:2} to (\ref{eq:PropConj0-1-1}), which only approximates the label-conditioned joint densities of (\ref{eq:PropConj0-1-1}). However, it is difficult to establish any results on the Kullback-Leibler
divergence for this more general class. }
\begin{prop}
\textcolor{black}{\label{prop:2-1}Given any labeled multi-object
density of the form (\ref{eq:PropConj0-1-1}) a $\delta$-GLMB which
preserves the cardinality distribution and the PHD of $\mathbf{\boldsymbol{\pi}}$
is given by
\begin{equation}
\boldsymbol{\hat{\pi}}(\mathbf{X})=\Delta({\mathbf{X}})\sum_{(c,I)\in{\mathbb{C}}\times{\mathcal{F}}({\mathbb{L}})}\delta_{I}({\mathcal{L}}({\mathbf{X}}))\hat{w}^{(c,I)}\left[\hat{p}^{(c,I)}\right]^{{\mathbf{X}}}\label{eq:ApproximateGLMB-1}
\end{equation}
where
\begin{flalign}
 \!\! \hat{w}^{(c,I)}&=w^{(c)}(I),\label{eq:MarginalizeGeneral-1-1-1}\\
 \!\! \hat{p}^{(c,I)}(x,\ell)&=1_{I}(\ell)p_{I-\{\ell\}}^{(c)}(x,\ell),\label{eq:MarginalizeGeneral-2-1-1}\\
 \!\! p_{\{\ell_{1},...,\ell_{n}\}}^{(c)}(x,\ell)&=\nonumber \\
 & \!\!\!\!\!\!\!\!\!\!\!\!\!\!\!\!\!\!\!\!\!\!\!\!\! \int \! p^{(c)}(\{(x,\ell),(x_{1},\ell_{1}),...,(x_{n},\ell_{n})\})d(x_{1},...,x_{n}).\label{eq:MarginalizeGeneral-3-1-1}
\end{flalign}}
\end{prop}
\vspace{-0.5cm}
\textcolor{black}{The proof follows along the
same lines as Proposition \ref{prop:2}.}
\begin{rem}
\textcolor{black}{Note that in \cite[Sec. V]{Beard14} a $\delta$-GLMB
was proposed to approximate a particular family of labeled RFS densities
that arises from multi-target filtering with merged measurements.
Our results show that the approximation used in \cite[Sec. V]{Beard14}
preserves the cardinality distribution and the PHD.}

In multi-object tracking, the matching of the cardinality distribution
and PHD in Proposition \ref{prop:2} is a stronger result than simply
matching the PHD alone. Notice that this property does not hold for
the LMB filter, as shown in \cite{Reuter2014} (Section III), due
to the imposed multi-Bernoulli parameterization of the cardinality
distribution.
\end{rem}

\section{Application to multi-target Tracking}

In this section we propose a multi-target tracking filter for generic
measurement models by applying the GLMB approximation result of Proposition
\ref{prop:2}. Specifically, we present the prediction and update
of the Bayes multi-target filter (\ref{eq:MTBayesUpdate})-(\ref{eq:MTBayesPred})
for the standard multi-target dynamic model as well as a generic measurement
model.

\subsection{Multi-target Filtering}

Following \cite{BTV13,Vo2014}, to ensure distinct \textcolor{black}{labels
we assign each target an o}rdered pair of integers $\ell=(k,i)$,
where $k$ is the time of birth and $i$ is a unique index to distinguish
targets born at the same time. The label space for targets born at
time $k$ is denoted by $\mathbb{L}_{k}$, and the label space for
targets at time $k$ (including those born prior to $k$) is denoted
as $\mathbb{L}_{0:k}$. Note that $\mathbb{L}_{k}$ and $\mathbb{L}_{0:k-1}$
are disjoint and $\mathbb{L}_{0:k}=\mathbb{L}_{0:k-1}\cup\mathbb{L}_{k}$.

A multi-target state $\mathbf{X}_{k}$ at time $k$, is a finite subset
of ${\mathcal{X=}}$ ${\mathbb{X}}{\mathcal{\times}}{\mathbb{L}}_{0:k}$.
Similar to the standard state space model, the multi-target system
model can be specified, for each time step $k$, via the \emph{multi-target
transition density} ${\bf f}_{k|k-1}(\cdot|\cdot)$ and the \emph{multi-target
likelihood function} $g_{k}(\cdot|\cdot)$, using the FISST notion
of integration/density. The \emph{multi-target posterior density}
(or simply multi-target posterior) contains all information on the
multi-target states given the measurement history. The multi-target
posterior recursion generalizes directly from the posterior recursion
for vector-valued states \cite{Doucet2000}, i.e. for $k\geq1$
\begin{flalign}
 &\!\!\!\! \mathbf{\boldsymbol{\pi}}_{0:k}(\mathbf{X}_{0:k}|z_{1:k})\propto\nonumber \\
 & \ \ g_{k}(z_{k}|\mathbf{X}_{k}){\bf f}{}_{k|k-1}(\mathbf{X}_{k}|\mathbf{X}_{k-1})\mathbf{\boldsymbol{\pi}}_{0:k-1}(\mathbf{X}_{0:k-1}|z_{1:k-1}),\label{eq:MTPosterior}
\end{flalign}
w\textcolor{black}{here $\mathbf{X}_{0:k}=(\mathbf{X}_{0},...,\mathbf{X}_{k})$
is the multi-target state history, and $z_{1:k}=(z_{1},...,z_{k})$
is the measurement history with $z_{k}$ denoting the measurement
at time $k$. Target trajectories or tracks are accommodated
in this formulation through the inclusion of a distinct label in the
target's state vector \cite{Goodman1997,Mahler2007,BTV13,Vu2014}.
The multi-target posterior (\ref{eq:MTPosterior}) then contains all
information on the random finite set of tracks, given the measurement
history.}

\textcolor{black}{In this work we are interested in the }\textcolor{black}{\emph{multi-target
filtering density}}\textcolor{black}{{} $\boldsymbol{\pi}_{k}$, a
marginal of the multi-target posterior, which can be propagated forward
recursively by the }\textcolor{black}{\emph{multi-target Bayes filter
}}\textcolor{black}{\cite{MahlerPHD2,Mahler2007}
\begin{align}
\boldsymbol{\pi}_{k}(\mathbf{X}_{k}|z_k) & =\frac{g_{k}(z_{k}|\mathbf{X}_{k})\boldsymbol{\pi}_{k|k-1}(\mathbf{X}_{k})}{\int g_{k}(z_{k}|\mathbf{X})\boldsymbol{\pi}_{k|k-1}(\mathbf{X})\delta\mathbf{X}},\label{eq:MTBayesUpdate}\\
\boldsymbol{\pi}_{k+1|k}(\mathbf{X}_{k+1}) & =\int{\bf f}{}_{k+1|k}(\mathbf{X}_{k+1}|\mathbf{X})\boldsymbol{\pi}_{k}(\mathbf{X}|z_k)\delta\mathbf{X},\label{eq:MTBayesPred}
\end{align}
where $\boldsymbol{\pi}_{k+1|k}$ is the }\textcolor{black}{\emph{multi-target
prediction density}}\textcolor{black}{{} to time $k+1$} (the dependence on the data is omitted for compactness). An analytic
solution to the multi-target Bayes filter for labeled states and track
estimation from the multi-target filtering density is given in \cite{BTV13}.
Note that a large volume of work in multi-target tracking is based
on filtering, and often the term \textquotedbl{}multi-target posterior\textquotedbl{}
is used in place of \textquotedbl{}multi-target filtering density\textquotedbl{}.
In this work we shall not distinguish between the filtering density
and the posterior density.

\subsection{Update}

In this section we apply the
proposed $\delta$-GLMB approximation to multi-target tracking
with a generic measurement model. We do not assume any particular structure for the multi-target
likelihood function $g(\cdot|\cdot)$ and hence the approach in this
section is applicable to any measurement model including point detections,
superpositional sensors and imprecise measurements \cite{Mahler2007,Papi2014}.
If the multi-target prediction density
is a $\delta$-GLMB of the form
\begin{equation}
\boldsymbol{\pi}_{k|k-1\!}({\mathbf{X}})=\Delta({\mathbf{X}})\!\!\!\!\sum\limits _{I\in{\mathcal{F}}({\mathbb{L}}_{0:k})}\!\!\!\delta_{I}({\mathcal{L}}({\mathbf{X}}))w_{k|k-1\!}^{(I)}\!\left[p_{k|k-1\!}^{(I)}\right]^{\!{\mathbf{X}}}\!,\label{eq:deltaGLMBprior}
\end{equation}
then the multi-target posterior density (\ref{eq:MTBayesUpdate}) becomes
\begin{equation}
\boldsymbol{\pi}_{k}(\mathbf{X}|z_{k})=\Delta(\mathbf{X})\!\!\!\!\!\sum_{I\in{\mathcal{F}}({\mathbb{L}}_{0:k})}\!\!\!\!\!\delta_{I\!}(\mathcal{L}(\mathbf{X}))w_{k}^{(I)}\!(z_{k})p_{k}^{(I)}(\mathbf{X}|z_{k}),\label{eq:JointPi}
\end{equation}
where
\begin{flalign}
\!\!\!\!\!  w_{k}^{(I)}(z_{k})& \propto w_{k|k-1}^{(I)}\eta_{z_{k}}(I),\label{eq:JointMixture-w}\\ 
\!\!\!\!\!  p_{k}^{(I)}(\mathbf{X}|z)&=g(z_{k}|\mathbf{X})[p_{k|k-1}^{(I)}]^{\mathbf{X}}/\eta_{z_{k}}(I),\label{eq:JointMixture-p}\\
\!\!\!\!\!  \eta_{z_{k}}(\{\ell_{1},...,\ell_{n}\})&=\int \! g(z_{k}|\{(x_{1},\ell_{1}),...,(x_{n},\ell_{n})\})\times\nonumber \\
 &\ \ \ \ \overset{n}{\underset{i=1}{\prod}}p_{k|k-1}^{(\{\ell_{1},...,\ell_{n}\})\!}(x_{i},\ell_{i})d(x_{1},...,x_{n}).\label{eq:JointMixture-n}
\end{flalign}
Note from (\ref{eq:JointMixture-p}) that after the update each multi-object exponential
$[p_{k|k-1}^{(I)}]^{{\mathbf{X}}}$ from the prior $\delta$-GLMB
becomes $p^{(I)}_{k}({\mathbf{X}}|z_{k})$, which is not necessarily a multi-object exponential. Hence, in general, (\ref{eq:JointPi}) is not a GLMB density.

\subsubsection{Separable Likelihood}

If targets are well separated in the measurement space, we can approximate the likelihood by a
separable one, i.e. $g(z_{k}|{\mathbf{X}})\approx\gamma_{z_{k}}^{{\mathbf{X}}}$,
and obtain an approximate GLMB posterior from Proposition 1:
\begin{equation}
\boldsymbol{\hat{\pi}}_{\!k}({\mathbf{X}}|z_{k})=\Delta({\mathbf{X}})\!\!\!\!\!\sum\limits _{I\in{\mathcal{F}}({\mathbb{L}}_{0:k})}\!\!\!\!\delta_{I\!}({\mathcal{L}}({\mathbf{X}}))\hat{w}_{k}^{(I)}\!(z_{k\!})\!\left[\hat{p}_{k}^{(I)\!}(\cdot|z_{k})\right]^{\!{\mathbf{X}}}\!\!\!\!,
\end{equation}
where
\begin{eqnarray}
\hat{w}_{k}^{(I)}(z_{k}) & \propto & w_{k|k-1}^{(I)}\left[\eta_{z_{k}}\right]^{I},\\
\hat{p}_{k}^{(I)}(x,\ell|z_{k}) & = & {p_{k|k-1}^{(I)}(x,\ell)\gamma_{z_{k}}(x,\ell)}/{\eta_{z_{k}}(\ell)},\\
\eta_{z_{k}}(\ell) & = & \left\langle p_{k|k-1}^{(I)}(\cdot,\ell),\gamma_{z_{k}}(\cdot,\ell)\right\rangle .
\end{eqnarray}

\subsubsection{General Case}

If instead targets are closely spaced, the separable likelihood assumption is violated, then it becomes necessary to directly approximate the multi-target posterior in (\ref{eq:JointPi}) which can be rewritten as:
\begin{equation}
\boldsymbol{{\pi}}_{k}({\mathbf{X}}|z_{k})={w}_{k}^{({\mathcal{L}}({\mathbf{X}}))}\!(z_{k})\Delta({\mathbf{X}}){p}_{k}^{({\mathcal{L}}({\mathbf{X}}))}({{\mathbf{X}}}|z_{k})
\end{equation}

\noindent It follows from Proposition \ref{prop:2}
that an approximate $\delta$-GLMB of the form (\ref{eq:ApproximateGLMB-1-1}), which matches the cardinality and PHD of the above multi-target posterior, as well as minimizing the Kullback-Leibler divergence from it, is given by
\begin{equation}
\boldsymbol{\hat{\pi}}_{\!k}({\mathbf{X}}|z_{k})=\Delta({\mathbf{X}})\!\!\!\!\!\sum\limits _{I\in{\mathcal{F}}({\mathbb{L}}_{0:k})}\!\!\!\!\delta_{I\!}({\mathcal{L}}({\mathbf{X}})){w}_{k}^{(I)}\!(z_{k\!})\!\left[\hat{p}_{k}^{(I)\!}(\cdot|z_{k})\right]^{\!{\mathbf{X}}}\!\!\!\!,
\end{equation}
where for each label set $I=\{\ell_{1},...,\ell_{n}\}$, the densities $\hat{p}_{k}^{(\{\ell_{1},...,\ell_{n}\})}(\cdot,\ell_{i}|z_{k})$, $i=1,...,n$ are the marginals of ${p}_{k}^{(\{\ell_{1},...,\ell_{n}\})}\{(\cdot,\ell_{1}),...,(\cdot,\ell_{n})\}|z_{k})$.
Notice that we retained the weights ${w}_{k}^{(I)}\!(z_k)$, given by 
(\ref{eq:JointMixture-w}), from the true posterior (\ref{eq:JointPi}).

\subsection{Prediction}

The standard multi-target dynamic model is described as follows. Given
the current multi-target state ${\mathbf{X}}^{\prime}$, each state
$(x^{\prime},\ell^{\prime})$ $\in{\mathbf{X}}^{\prime}$ either continues
to exist at the next time step with probability $p_{S}(x^{\prime},\ell^{\prime})$
and evolves to a new state $(x,\ell)$ with probability density $f_{k+1|k}(x|x^{\prime},\ell^{\prime})\delta_{\ell}(\ell^{\prime})$,
or dies with probability $1-p_{S}(x^{\prime},\ell^{\prime})$. The multi-target state at the next time is the superposition
of surviving and new born targets. The
set of new targets born at the next time step is distributed according
to a birth density ${\mathbf{f}}_{B}$ on ${\mathcal{F}}({\mathbb{X}}\times{\mathbb{L}}_{k+1})$, given by
\begin{equation}
{\mathbf{f}}_{B}({\mathbf{Y}})=\Delta({\mathbf{Y}})w_{B}({\mathcal{L}}({\mathbf{Y}}))\left[p_{B}\right]^{{\mathbf{Y}}}\label{eq:Birth_transition}
\end{equation}
This birth model covers labeled Poisson,
labeled IID cluster and LMB. We use an LMB
birth model with 
\begin{eqnarray}
w_{B}(L) & = & \prod\limits _{i\in{\mathbb{L}}_{k}}\left(1-r_{B}^{(i)}\right)\prod\limits _{\ell\in L}\frac{1_{{\mathbb{L}}_{k}}(\ell)r_{B}^{(\ell)}}{1-r_{B}^{(\ell)}},\\
p_{B}(x,\ell) & = & p_{B}^{(\ell)}(x).
\end{eqnarray}

Following \cite{BTV13}, if the current multi-target posterior has the following $\delta$-GLMB form
\begin{equation}
\boldsymbol{\pi}_{k}(\mathbf{X})=\Delta(\mathbf{X})\!\!\sum_{I\in{\mathcal{F}}({\mathbb{L}}_{0:k})}\!\!\delta_{I}(\mathcal{L}(\mathbf{X}))w_{k}^{(I)}\left[p_{k}^{(I)}\right]^{\mathbf{X}}\!\!,
\end{equation}
then the multi-target prediction (\ref{eq:MTBayesPred}) is also a $\delta$-GLMB:
\begin{equation}
{\mathbf{\boldsymbol{\pi}}}_{\!k+1|k}({\mathbf{X}})=\Delta({\mathbf{X}})\!\!\!\!\!\sum_{I\in\mathcal{F}(\mathbb{L}_{0:k+1})}\!\!\!\!\!\delta_{I\!}(\mathcal{L}(\mathbf{X}))w_{k+1|k}^{(I)}\!\left[p_{k+1|k}^{(I)}\right]^{\!{\mathbf{X}}}
\end{equation}
where
\begin{flalign}
 & w_{k+1|k}^{(I)}=w_{S}^{(I)}(I\cap\mathbb{L}_{0:k})w_{B}(I\cap\mathbb{L}_{k+1}),\nonumber\\
 & w_{S}^{(I)}(L)=[\eta_{S}^{(I)}]^{L}\sum_{J\subseteq\mathbb{L}_{0:k}}1_{J}(L)[1-\eta_{S}^{(I)}]^{J-L}w_{k}^{(I)}(J),\nonumber\\
 & p_{k+1|k}^{(I)}(x,\ell)=1_{\mathbb{L}_{0:k}}(\ell)p_{S}^{(I)}(x,\ell)+(1-1_{\mathbb{L}_{0:k}}(\ell))p_{B}(x,\ell),\nonumber\\
 & p_{S}^{(I)}(x,\ell)=\frac{\left\langle p_{S}(\cdot,\ell)f_{k+1|k}(x|\cdot,\ell),p_{k}^{(I)}(\cdot,\ell)\right\rangle }{\eta_{S}^{(I)}(\ell)},\nonumber\\
 & \eta_{S}^{(I)}(\ell)=\left\langle p_{S}(\cdot,\ell),p_{k}^{(I)}(\cdot,\ell)\right\rangle.\nonumber
\end{flalign}
\textcolor{black}The above {Eqs. explicitly
describe the calculation of the parameters of the predicted multi-target
density from the parameters of the previous multi-target density \cite{Vo2014}.}

\section{Numerical Results}

In this section we verify the proposed GLMB approximation technique
via an application to recursive multi-target tracking with radar power
measurements. Target tracking is usually performed on data that have
been preprocessed into point measurements or detections. The bulk
of multi-target tracking algorithms in the literature are designed
for this type of data \cite{Blackman1999,Bar2011,Mahler2007,Mallick2012}.
Compressing information from the raw measurement into a finite set
of points is very effective for a wide range of applications. However,
for applications with low SNR, this approach may not be adequate as
the information loss incurred in the compression becomes significant.
Consequently, it becomes necessary to make use of all information
contained in the pre-detection measurements, which in turn requires
more advanced sensor models and algorithms.

We first describe the single-target dynamic model and multi-target
measurement equation used to simulate the radar power measurements.
We then report numerical results for the separable likelihood approximation
and GLMB posterior approximation. Throughout this section our recursive
multi-target tracker is implemented with a particle filter approximation
\cite{Doucet2000,Gor93} of the GLMB density given in \cite{Vo2014}.

\subsection{Dynamic Model}

The kinematic part of the single-target state  ${\mathbf{x}}_{k}=(x_{k},\ell_{k})$
at time $k$ comprises the planar position, velocity vectors 
in 2D Cartesian coordinates, and the unknown modulus
of the target complex amplitude $\zeta_{k}$, respectively, i.e. ${x}_{k}=[p_{x,k},\dot{p}_{x,k},p_{y,k},\dot{p}_{y,k},\zeta_{k}]^{T}$. A Nearly
Constant Velocity (NCV) model is used to describe the target dynamics,
while a zero-mean Gaussian random walk is used to model the fluctuations
of the target complex amplitude, i.e.
\[
x_{k+1}=Fx_{k}+v_{k},~~v_{k}\sim{\mathcal{N}}\left(0;Q\right)
\]
where $F  =\mbox{diag}(F_{1},F_{1},1)$, $Q  =\mbox{diag}(qQ_{1},qQ_{1},a_{\zeta}T_{s})$, 
\begin{align*}
& F_{1}=\left[\begin{array}{cc}
1 & T_{s}\\
0 & 1
\end{array}\right], ~~Q_{1} =\left[\begin{array}{cc}
\frac{T_{s}^{3}}{3} & \frac{T_{s}^{2}}{2}\\
\frac{T_{s}^{2}}{2} & T_{s}
\end{array}\right]
\end{align*}
with $T_{s}$, $q$, and $a_{\zeta}$ denoting the radar
sampling time,  the power spectral density of the process noise,
and the amplitude fluctuation in linear domain, respectively.

\subsection{TBD Measurement Equation}

A target ${\mathbf{x}}\in{\mathbf{X}}$ illuminates a set of cells
$C({\mathbf{x}})$, usually referred to
as the \textit{target template}. A radar positioned at the Cartesian
origin collects a vector measurement $z=[z^{(1)},..., z^{(m)}]$
consisting of the power signal returns $z^{(i)}=|z_{A}^{(i)}| ^{2}$, where 
\begin{equation}
z_{A}^{(i)}=\sum_{{\mathbf{x\in X}}}1_{C({\mathbf{x}})}(i)A({\mathbf{x}})h_{A}^{(i)}({\mathbf{x}})+w^{(i)} \nonumber
\end{equation}
is the complex signal in cell $i$, with:
\begin{itemize}
\item $w^{(i)}$ denoting zero-mean white circularly symmetric complex Gaussian noise with variance
$2\sigma_{w}^{2}$;
\item $h_{A}^{(i)}({\mathbf{x}})$ denoting the point spread function value in cell
$i$ from a target with state ${\mathbf{x}}${\footnotesize{}
\begin{align}
h_{A}^{(i)}({\mathbf{x}})=\exp\left(-{\displaystyle \frac{(r_{i}-r({\mathbf{x}}))^{2}}{2R}-{\displaystyle \frac{(d_{i}-d({\mathbf{x}}))^{2}}{2D}-{\displaystyle \frac{(b_{i}-b({\mathbf{x}}))^{2}}{2B}}}}\right)\nonumber
\end{align}
}where $R$, $D$, $B$ are resolutions for range, Doppler, bearing;
$r({\mathbf{x}})=\sqrt{p_{x}^{2}+p_{y}^{2}}$, $d({\mathbf{x}})=-(\dot{p}_{x}p_{x}+\dot{p}_{y}p_{y})/r({\mathbf{x}})$,
$b({\mathbf{x}})=\mbox{atan2}(p_{y},p_{x})$ are range, Doppler,
bearing, given the target state $\mathbf{x}$; and $r_{i},d_{i},b_{i}$
are cell centroids;
\item $A({\mathbf{x}})$ denoting the complex echo of target ${\mathbf{x}}$,
which for a Swerling $0$ model is constant in modulus
\begin{equation}
A({\mathbf{x}})=\bar{A}e^{j\theta},\ \ \theta\sim{\mathcal{U}}_{\lbrack0,2\pi)}. \nonumber
\end{equation}

\end{itemize}
\noindent Let $\hat{z}^{(i)} = |\hat{z}_{A}^{(i)}|^{2}$ \textcolor{black}{be the noiseless power return in
cell $i$, where}
\begin{equation}
\hat{z}_{A}^{(i)} = \sum_{{\mathbf{x}}\in{\mathbf{X}}}1_{C({\mathbf{x}})}(i)\bar{A}h_{A}^{(i)}({\mathbf{x}}). \nonumber
\end{equation}
\noindent The measurement $z^{(i)}$ in each cell  follows a non-central
chi-squared distribution with $2$ degrees of freedom and non-centrality
parameter $\hat{z}_{A}^{(i)}$, and simplifies to a central chi-squared
distribution with $2$ degrees of freedom when $\hat{z}_{A}^{(i)}=0$.
Consequently, the likelihood ratio for cell $(i)$ is given by:
\begin{equation}
\ell(z^{(i)}|{\mathbf{X}})=\exp\left(-0.5\hat{z}^{(i)}\right)I_{0}\left(\sqrt{z^{(i)}\hat{z}^{(i)}}\right)\label{eq:likelihood}
\end{equation}
\noindent where $I_{0}(\cdot)$ is the modified Bessel function, which can be evaluated
using the approximation given in \cite{Bessel}. 

Given a vector measurement $z$ the likelihood function of the multi-target state $\mathbf{X}$ takes the form \textcolor{black}{
\begin{align}
g(z|{\mathbf{X}}) & \propto\prod_{i\in{\displaystyle {\displaystyle \cup_{\mathbf{x}\in\mathbf{X}}}}C(\mathbf{x})}\ell(z^{(i)}|{\mathbf{X}}),\label{eq:like_ratio}
\end{align}
} 
\noindent Notice that eqs.
(\ref{eq:likelihood})-(\ref{eq:like_ratio}) capture the superpositional
nature of the power returns for each measurement bin due to the possibility
of closely spaced targets target, i.e. overlapping target templates.
The separable likelihood assumption is obtained from eqs. (\ref{eq:likelihood})-(\ref{eq:like_ratio})
by assuming that at most one target contributes to the power return
from each cell $(i)$,
\noindent
\begin{eqnarray*}
\hat{z}^{(i)} & = & |\hat{z}_{A}^{(i)}|^{2}=\begin{cases}
|\bar{A}h_{A}^{(i)}({\mathbf{x}})|^{2}, & \exists\mathbf{x}\in\mathbf{X}:i\in C({\mathbf{x}})\\
0, & \mbox{otherwise}
\end{cases}
\end{eqnarray*}

In the numerical examples we use $10\log{\left(\bar{A}^{2}/({2\sigma_{w}^{2}})\right)}$ as the signal-to-noise
ratio (SNR) definition, 
and choosing $\sigma_{w}^{2}=1$ implies $\bar{A}=\sqrt{2\cdot10^{SNR/10}}$.
\vspace{-0.3cm}
\begin{table}[htbp]
\protect\protect\caption{Common Parameters used in Simulations}
\centering{}\label{tab:Tab}%
\begin{tabular}{|c|c|c|}
\hline
Parameter  & Symbol  & Value\tabularnewline
\hline
\hline
Signal-to-Noise Ratio  & $\mbox{SNR}$  & $7\mbox{dB}$\tabularnewline
\hline
Power Spectral Density  & $q$  & $3\mbox{m}^{2}/\mbox{s}^{3}$\tabularnewline
\hline
Amplitude Fluctuation  & $a_{\rho}$  & $1$\tabularnewline
\hline
$1^{st}$ Birth Point Coordinates  & $\mathbf{x}_{B}^{1}$  & $\left[1250,-10,1000,-10\right]$\tabularnewline
\hline
$2^{nd}$ Birth Point Coordinates  & $\mathbf{x}_{B}^{2}$  & $\left[1000,-10,1250,-10\right]$\tabularnewline
\hline
$3^{rd}$ Birth Point Coordinates  & $\mathbf{x}_{B}^{3}$  & $\left[1250,-10,1250,-10\right]$\tabularnewline
\hline
Birth Probability  & $P_{B}$  & $0.01$\tabularnewline
\hline
Survival Probability  & $P_{S}$  & $0.99$\tabularnewline
\hline
$n^{\circ}$of particles per target  & $N_{p}$  & $1000$\tabularnewline
\hline
\end{tabular}
\end{table}
\vspace{-0.3cm}
\begin{table}[htbp]
\protect\protect\caption{Separable Likelihood Parameters}
\centering{}\label{tab:Tab-sep}%
\begin{tabular}{|c|c|c|}
\hline
Parameter  & Symbol  & Value\tabularnewline
\hline
\hline
Range Resolution  & $R$  & $5\mbox{m}$\tabularnewline
\hline
Azimuth Resolution  & $B$  & $1{}^{\circ}$\tabularnewline
\hline
Doppler Resolution  & $D$  & $1\mbox{m}/\mbox{s}$\tabularnewline
\hline
Sampling Time  & $T_{s}$  & $2\mbox{s}$\tabularnewline
\hline
Birth Covariance & $Q_{B}$ & $\mbox{diag}\left([25,4,25,4]\right)$\tabularnewline
\hline
\end{tabular}
\end{table}

\subsection{Separable Likelihood Results}

In this section we report simulation results for a radar TBD scenario
under the separable likelihood assumption, which
is valid when targets do not overlap at any time.
This implies that the birth density is relatively informative compared
to the targets kinematics. This apparently obvious requirement is
necessary to avoid a bias in the estimated number of targets due to
new target or birth hypotheses which always violate the separable
likelihood assumption.

\begin{figure}
\vspace{-0.4cm}
\centering{}\includegraphics[width=8cm]{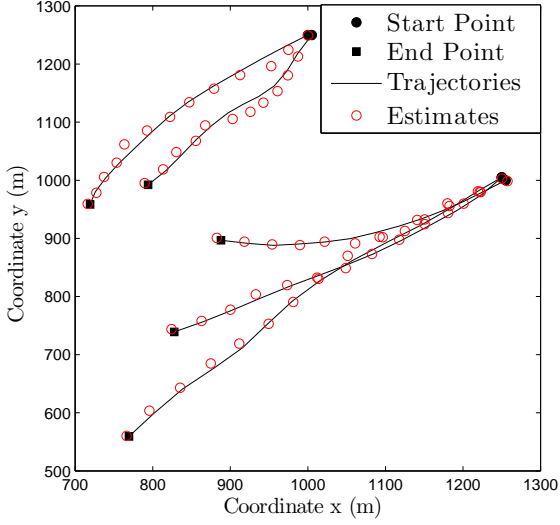}
\vspace{-0.3cm}
\caption{Separable likelihood scenario. Targets appear from the top right corner
and move closer to the radar positioned at the Cartesian origin.}
\label{fig:scenario1}
\end{figure}

\begin{figure}
\begin{centering}
\includegraphics[width=8cm]{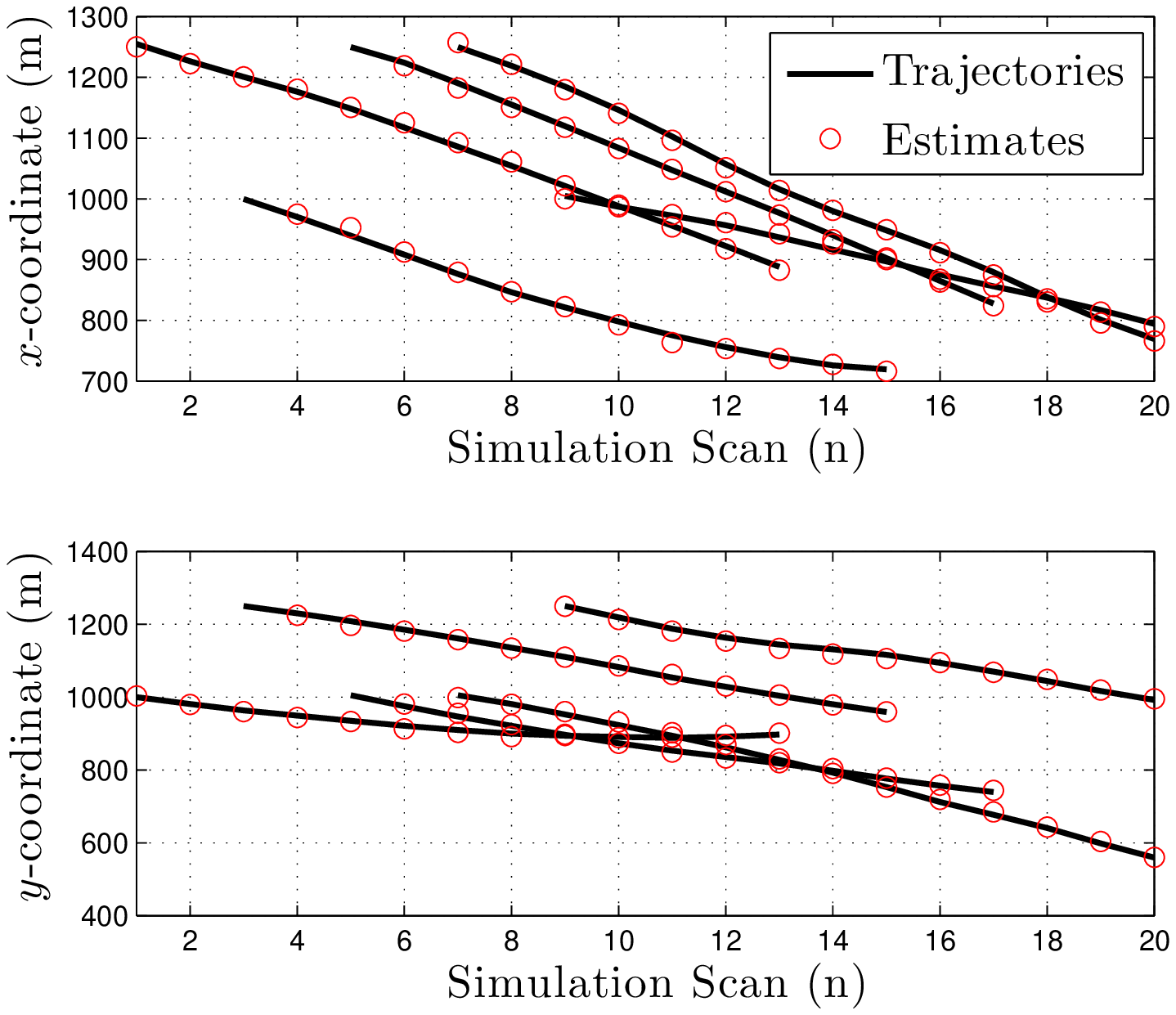}
\par\end{centering}
\vspace{-0.1cm}
\protect\protect\caption{Separable likelihood scenario. Estimated trajectories along the $x$
and $y$ coordinates.}
\label{fig:1_xy}
\end{figure}

\begin{figure}
\begin{centering}
\includegraphics[width=8cm]{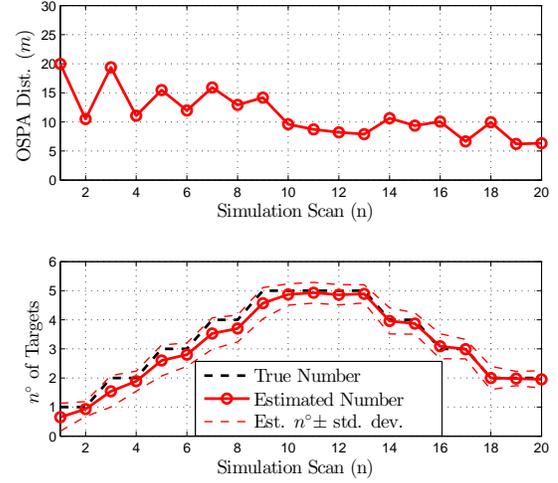}
\par\end{centering}
\vspace{-0.1cm}
\protect\protect\caption{Separable likelihood scenario. Monte Carlo results for estimated number
of targets (top) and the OSPA distance (bottom) with cut-off $c=50\mbox{m}$}
\label{fig:1_ospa}
\end{figure}

The considered scenario is depicted in Fig. \ref{fig:scenario1}:
we have a time varying number of targets due to various births and
deaths with a maximum of $5$ targets present mid scenario. The parameters
are reported in Tables \ref{tab:Tab} and \ref{tab:Tab-sep}. Fig.
\ref{fig:1_xy} shows the estimation results for a single trial
along the $x$ and $y$ coordinates, and Fig. \ref{fig:1_ospa}
shows the Monte Carlo results for the estimated number of targets
and positional OSPA distance. Notice that the average estimated number
of targets slightly differs from the true number due to closely spaced
targets (see Fig. \ref{fig:scenario1}), but the overall performance
is satisfactory given the low $\mbox{SNR}$ of $7\mbox{dB}$.

\subsection{Non-Separable Likelihood Results}

\begin{figure}[htbp]
\vspace{-0.6cm}
\begin{centering}
\includegraphics[width=8cm]{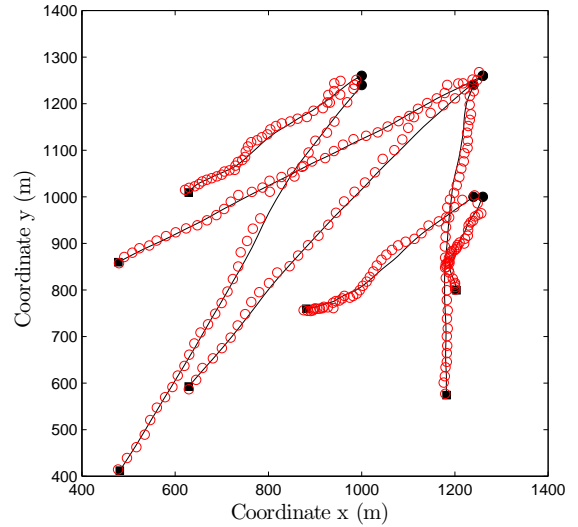}
\par\end{centering}
\vspace{-0.2cm}
\protect\protect\caption{Non-separable likelihood scenario. Targets appear from the top right
corner and move closer to the radar positioned at the Cartesian origin.}
\label{fig:scenario2}
\end{figure}

\begin{figure}[tp]
\vspace{-0.9cm}
\begin{centering}
\includegraphics[width=8cm]{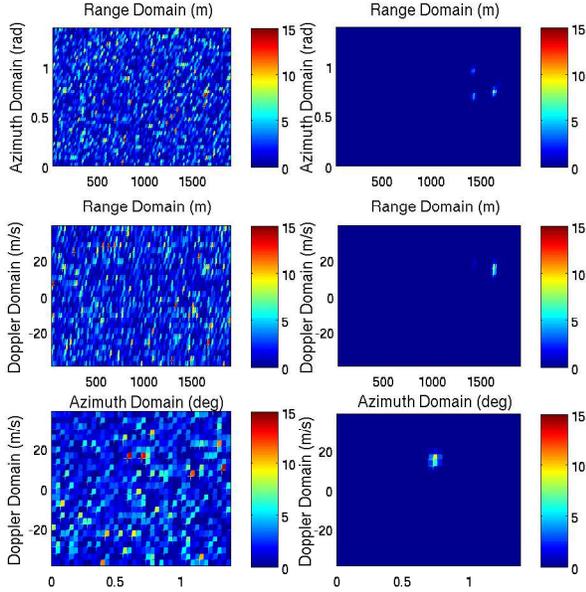}
\par\end{centering}
\protect\caption{Non-separable likelihood scenario. Range-Azimuth, Range-Doppler, and
Azimuth-Doppler maps at time instant $k=19$. Ideal or noiseless measurement
(\emph{right column}), and noisy measurement (\emph{left column}). Notice that for
each 2D map, the index of the $3{}^{rd}$ coordinate is such that
all maps refer to the same group of targets. Specifically, the target
reflection around ($1500$m,$0.8^{\circ}$,$18$m/s) is due to two
targets in the same Radar cell.}
\label{fig:Measurement_joint}
\end{figure}

In this section we consider a more difficult radar TBD scenario where
the separable likelihood assumption would lead to a bias on the estimated
number of targets. Fig. \ref{fig:scenario2} shows a time varying
number of targets due to various births and deaths with a maximum
of $7$ targets present mid scenario. Fig. \ref{fig:Measurement_joint}
shows range-azimuth, range-Doppler, and azimuth-Doppler maps of the
received power returns. Notice that for each 2D map, the index of
the $3{}^{rd}$ coordinate is such that all maps refer to the same
group of targets. Specifically, the target reflection around ($1500$m,
$0.8^{\circ}$, $18$m/s) is due to two targets in the same Radar
cell. This leads to the so-called unresolved target problem, which
usually results in track loss when using a standard detection based
approach or a separable likelihood assumption. The parameters used
in simulation are reported in Tables \ref{tab:Tab} and \ref{tab:Tab-gen}.
\begin{table}[htbp]
\protect\protect\caption{Non-Separable Likelihood Parameters}
\centering{}\label{tab:Tab-gen}%
\begin{tabular}{|c|c|c|}
\hline
Parameter  & Symbol  & Value\tabularnewline
\hline
\hline
Range Resolution  & $R$  & $20\mbox{m}$\tabularnewline
\hline
Azimuth Resolution  & $B$  & $2{}^{\circ}$\tabularnewline
\hline
Doppler Resolution  & $D$  & $2\mbox{m}/\mbox{s}$\tabularnewline
\hline
Sampling Time  & $T_{s}$  & $1\mbox{s}$\tabularnewline
\hline
Birth Covariance & $Q_{B}$ & $\mbox{diag}\left([400,100,400,100]\right)$\tabularnewline
\hline
\end{tabular}
\end{table}

\begin{figure}[htbp]
\begin{centering}
\includegraphics[width=8cm]{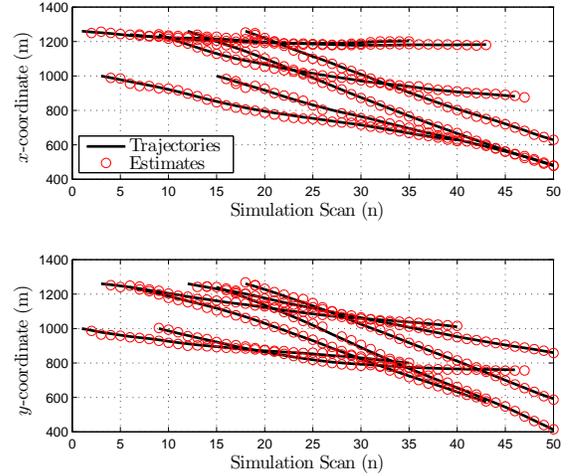}
\par\end{centering}
\protect\protect\caption{Non-separable likelihood scenario.Estimated trajectories along the
$x$ and $y$ coordinates.}
\label{fig:2_xy}
\end{figure}

\begin{figure}[tbph]
\begin{centering}
\includegraphics[width=8cm]{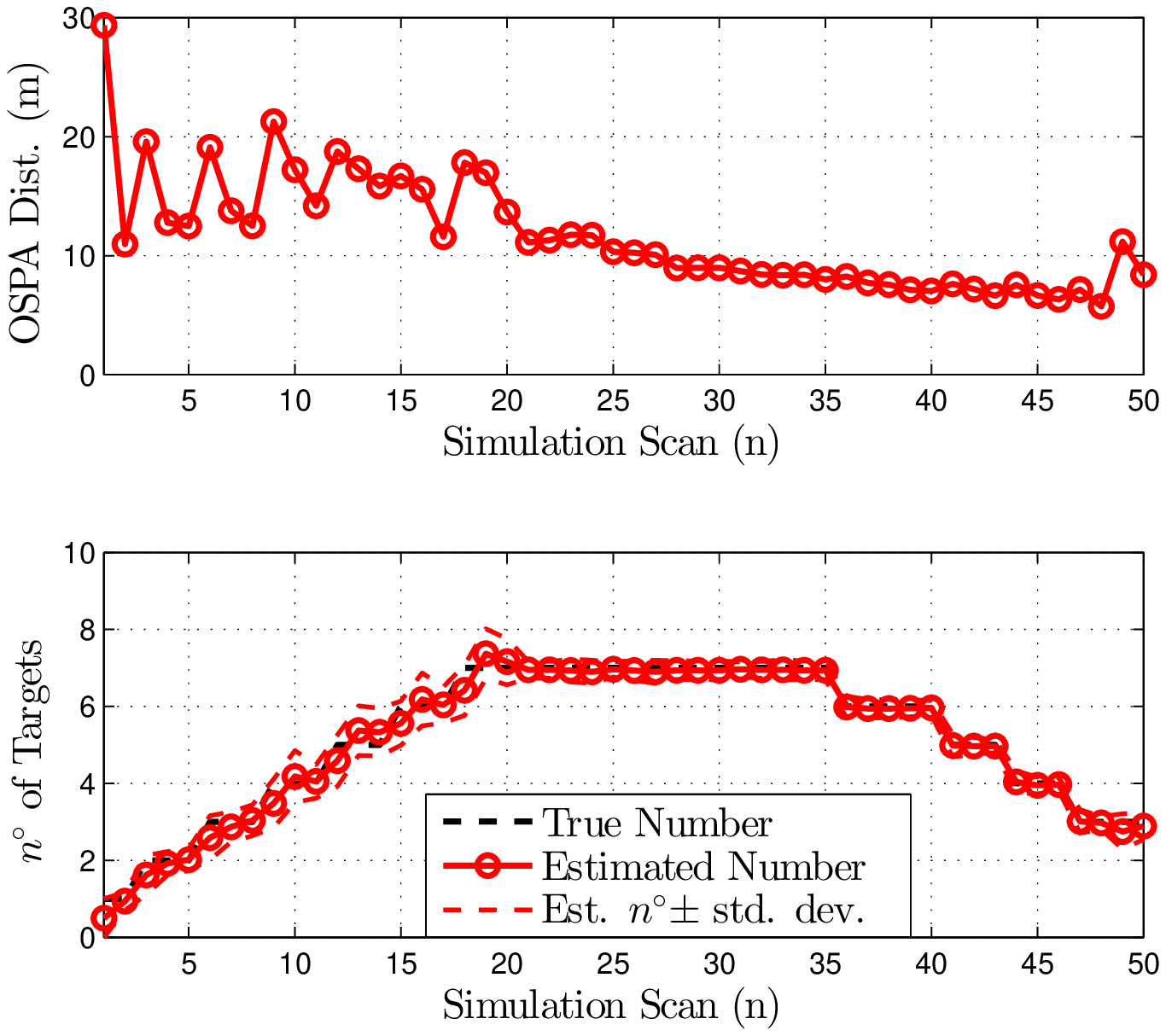}
\par\end{centering}
\protect\protect\caption{Non-separable likelihood scenario. Monte Carlo results for estimated
number of targets (top) and the OSPA distance (bottom) with cut-off
$c=50\mbox{m}$.}
\label{fig:2_ospa}
\end{figure}

The estimation results for a single trial along the $x$
and $y$ coordinates are shown in Fig. \ref{fig:2_xy}, and the Monte Carlo
results for the estimated number of targets and positional OSPA error is shown
in Fig. \ref{fig:2_ospa}. The results demonstrate that the proposed
GLMB approximation exhibits satisfactory tracking performance.

\section{Conclusions}

This paper has proposed a tractable class of GLMB approximations for
labeled RFS densities. \textcolor{black}{In particular, we derived
from this class of GLMBs an approximation that can capture the statistical
dependence between targets, preserves the cardinality distribution
and the PHD, as well as minimizes the Kullback-Leibler divergence.}
The result has particular significance in multi-target tracking since
it leads to tractable recursive filter implementations with formal
track estimates for a wide range of non-standard measurement models.
A radar based TBD example with low SNR and a time varying number of
closely spaced targets was presented to verify the theoretical result.
The key result presented in Section \ref{sec:GLMBapprox} is not only
important to recursive multi-target filtering but is also generally
applicable to statistical estimation problems involving point processes
or random finite sets.

\bibliographystyle{IEEEtran}


\end{document}